\documentclass[sigconf,noacm]{acmart}

\usepackage{booktabs}
\usepackage{tabularx}
\usepackage{array}

\AtBeginDocument{%
  }

\setcopyright{none} 
\renewcommand\footnotetextcopyrightpermission[1]{}
\begin{document}

\title[From Assistance to Agency]{From Assistance to Agency: Rethinking Autonomy and Control in CI/CD Pipelines}

\author{Marcus Emmanuel Barnes}
\orcid{0009-0000-7571-9085}
\affiliation{%
  \department{Faculty of Information}  
  \institution{University of Toronto}
  \city{Toronto}
  \state{Ontario}
  \country{Canada}
}
\email{marcus.barnes@utoronto.ca}

\author{Taher A. Ghaleb}
\orcid{0000-0001-9336-7298}
\affiliation{%
  \department{Department of Computer Science} 
  \institution{Trent University}
  \city{Peterborough}
  \state{Ontario}
  \country{Canada}}
\email{taherghaleb@trentu.ca}

\author{Safwat Hassan}
\orcid{0000-0001-7090-0475}
\affiliation{%
  \department{Faculty of Information}
  \institution{University of Toronto}
  \city{Toronto}
  \state{Ontario}
  \country{Canada}
}
\email{safwat.hassan@utoronto.ca}


\begin{abstract}
AI agents are assuming active roles in Continuous Integration and Continuous Deployment (CI/CD) workflows, yet the research community lacks a shared vocabulary for describing what it means for CI/CD to be \emph{agentic}, how much decision authority is delegated, and where control should reside. This paper presents a vision of agentic CI/CD in which the central challenge is not improving task performance but designing \emph{authority transfer}, defined as the delegation of operational decisions from human-controlled pipelines to agent systems under specified constraints and recourse mechanisms.

To structure this argument, we introduce a distinction between \emph{data-plane authority} (localized interventions such as patch generation and test reruns) and \emph{control-plane authority} (modifications to pipeline configuration, deployment policies, and approval gates). Drawing on research prototypes and industrial platforms, we show that current systems operate mainly at the data plane under \emph{bounded autonomy}, with safety achieved through surrounding governance infrastructure rather than intrinsic agent guarantees. We identify three recurring patterns: constrained autonomy as the dominant design, external governance as the primary safety mechanism, and a widening gap between deployment momentum and evaluation methodology. We propose a research agenda in which control-plane safety and governance mechanisms represent the most urgent open problem, followed by formalization of autonomy boundaries, evaluation frameworks, and human--agent coordination.
\end{abstract}



\makeatletter
\renewcommand{\shortauthors}{}%
\renewcommand{\shorttitle}{}%
\let\@acmConference\@empty
\let\@acmYear\@empty
\let\@acmISBN\@empty
\let\@acmDOI\@empty
\let\@acmPrice\@empty

\fancypagestyle{emptyheadstyle}{%
  \fancyhf{}
  \fancyhead[L]{}
  \fancyhead[C]{}
  \fancyhead[R]{}
  \fancyfoot[C]{\thepage}
  \renewcommand{\headrulewidth}{0pt}%
  \renewcommand{\footrulewidth}{0pt}%
}
\pagestyle{emptyheadstyle}
\thispagestyle{emptyheadstyle}

\def\@copyrightspace{}%
\makeatother

\maketitle

\section{Introduction}

Continuous Integration and Continuous Deployment (CI/CD) is central to modern software delivery \cite{Shahin2017CIDCDAccess, FitzgeraldStol2017ContinuousSE}. AI techniques have long supported tasks such as failure diagnosis and operational monitoring \cite{MishraOtaiwi2020DevOpsSLR, Notaro2021AIOpsSurvey}, typically as predictive or advisory components embedded within specific pipeline stages. In these systems, decision authority remains with human operators or fixed pipeline logic.

Recent advances in large language model-based systems and software engineering agents \cite{Hou2024LLMsForSESLR, LiuEtAl2024LLMAgentsSESurvey, WangEtAl2024AgentsInSESurvey} have introduced components capable of planning, tool use, and action execution. Industrial platforms now describe agents that observe pipeline state, generate patches or configuration changes, and submit modifications through repository workflows \cite{GitHubAgenticCI2026, GitLabDuoFixPipelineDocs, AmazonQGitHubAgentsDocs, RunGeminiCliGitHubAction, NxSelfHealingCIDocs}. Despite this rapid adoption, the research community lacks a shared vocabulary for characterizing how autonomous these systems are and where decision authority resides, making it difficult to compare systems labeled as “agentic” but with fundamentally different human-override and governance mechanisms.

We argue that the central challenge in agentic CI/CD is not improving remediation performance but designing \emph{authority transfer}, defined as the delegation of operational decisions from human-controlled pipelines to agent systems under specified constraints and recourse mechanisms. To structure this problem, we distinguish between \emph{data-plane authority}—localized interventions within a pipeline execution (e.g., patch generation, test reruns)—and \emph{control-plane authority}—modifications to pipeline configuration, deployment policies, and approval gates. Delegating control-plane authority alters organizational risk boundaries, yet current systems overwhelmingly confine agents to the data plane.

Drawing on research prototypes and industrial platforms, we characterize the current state as bounded autonomy~\cite{HuebscherMcCann2008AutonomicSurvey}, where agents operate within constrained action scopes and decision authority remains mediated by governance mechanisms such as human approval workflows. Across these systems, three recurring patterns emerge. First, constrained autonomy dominates: agents operate within predefined action spaces, with human approval mediating integration. Second, safety is achieved primarily through surrounding governance infrastructure (e.g., pull requests and policy gates) rather than intrinsic agent guarantees. Third, deployment momentum is outpacing evaluation methodology, and the field lacks benchmarks suited to decision-making agents.

The contribution of this paper is to reframe agentic CI/CD as a problem of authority transfer and to introduce the data-plane/control-plane distinction as a vocabulary for reasoning about autonomy and governance, with implications for system design and evaluation.

We conclude by outlining a research agenda that prioritizes control-plane safety and governance mechanisms, followed by formalization of autonomy boundaries, evaluation frameworks, and human--agent coordination. Treating authority transfer as a first-class design problem enables more precise reasoning about agentic behavior in CI/CD pipelines.

\section{Agentic CI/CD in Practice: Examples from Research and Industry}

To ground the discussion, we draw on research and industrial documentation. Because many developments originate from commercial platforms, both sources help clarify how agentic capabilities are framed and constrained in practice. Our goal is not exhaustive coverage, but a capability-focused set of exemplars that motivate the observations and research agenda that follow; this paper is not a systematic literature review but a synthesis of representative research prototypes and publicly documented industrial systems.

\subsection{Research Prototypes}

Early examples of agentic behavior in CI/CD appear in automated repair systems. Repairnator continuously monitors CI failures, synthesizes candidate patches, and submits pull requests for human review \cite{MonperrusEtAl2019RepairnatorUbiquity, UrliYuSeinturierMonperrus2018RepairBotDesign}. R-HERO extends this pattern by incorporating continual learning into patch generation \cite{BaudryEtAl2021SoftwareRepairRobot}. However, integration authority remains mediated by pull request workflows, limiting their autonomy.

More recent LLM-based repair systems exhibit increasing agent-like behavior. RepairAgent~\cite{RepairAgent2025} autonomously plans and executes multi-step repair strategies by invoking tools, gathering diagnostic information, and iterating on candidate patches. ChatRepair~\cite{ChatRepair2024} uses conversational LLM interaction to fix bugs through iterative dialogue with test feedback. These systems demonstrate richer planning and tool-use capabilities, yet human reviewers retain final say over merging. BuildSheriff~\cite{ZhangBuildSheriff2022} addresses a complementary problem, triaging CI test failures to their root causes, illustrating how diagnostic intelligence can precede and support remediation agents.

LogSage analyzes CI/CD logs to detect failures and support remediation actions in an industrial setting \cite{LogSageArxiv2506_03691}. Adjacent DevSecOps research such as AutoGuard frames self-healing as a proactive control layer that can adapt pipeline responses using reinforcement learning \cite{anugula2025autoguard}. Bouzenia and Pradel~\cite{Bouzenia2025} present an LLM agent capable of autonomously setting up and executing test suites for arbitrary projects, demonstrating that agentic capabilities are extending beyond patch generation toward broader build and test automation.

A notable gap separates agent behavior at the \emph{development layer} from the \emph{delivery layer}. Li et al.\ present AIDev, a large-scale study of AI agent activity on GitHub repositories, enabling analysis of agent-authored pull requests and interaction patterns with human contributors~\cite{AIDev}. While this work offers empirical foundations for understanding how agents contribute code, it does not address integration, deployment, or control-plane authority within CI/CD pipelines. This gap is significant: empirical evidence exists for code contribution, but not for release governance.

\subsection{Industrial Systems}

Industrial platforms increasingly describe CI/CD functionality using agent-oriented language, with consistent patterns across vendors. Major platforms (e.g., GitHub, GitLab, Amazon Q, and Google) now integrate agentic CI/CD capabilities while preserving human-mediated merge authority~\cite{GitHubAgenticCI2026, GitHubCopilotCodingAgent, GitLabDuoFixPipelineDocs, AmazonQGitHubAgentsDocs, RunGeminiCliGitHubAction}.

Nx Self-Healing CI, Datadog's Bits AI Dev Agent, and Gitar's autonomous build-fix engine each monitor pipeline outcomes, generate remediation actions, and route proposed changes through pull request mechanisms~\cite{NxSelfHealingCIDocs, DatadogBitsAIDevAgent, GitarAutonomousBuildFixes}. Dagger has published architectural blueprints for constructing self-healing pipelines using AI agents within containerized CI environments~\cite{DaggerSelfHealingPipelines}.

These systems exhibit properties associated with agency: perception of pipeline signals, decision-making over candidate actions, and execution through artifact generation. However, applying the data-plane and control-plane distinction, we observe that all publicly documented systems concentrate agent behavior at the data plane. Continuous delivery research emphasizes the importance of release governance and deployment discipline \cite{Shahin2017CIDCDAccess, FitzgeraldStol2017ContinuousSE, Humble2010ContinuousDelivery}. No system we examined delegates control-plane authority (i.e., modification of deployment policies, approval gates, rollback thresholds, or workflow definitions) without human approval, consistent with recent empirical evidence showing that agent interactions with CI/CD configurations remain limited in frequency and vary in success~\cite{ghaleb2026ai}.

\subsection{Architectural Implications}
The bounded autonomy observed across current systems has architectural consequences. CI/CD pipelines are traditionally deterministic execution graphs governed by explicit approval gates and disciplined deployment strategies~\cite{FitzgeraldStol2017ContinuousSE,Shahin2017CIDCDAccess}. Introducing agentic components embeds adaptive decision loops within these structures.

These structures resemble feedback-loop architectures from autonomic systems research, particularly the MAPE-K (Monitor-Analyze-Plan-Execute over shared Knowledge) reference model~\cite{Kephart2003AutonomicVision, WhiteHWCK2004ICAC}. However, the analogy is limited in important ways. MAPE-K assumes a single managed element with a well-defined adaptation policy, and its control loop is typically designed by the system architect with full authority over the feedback mechanism~\cite{Cheng2009SelfAdaptiveRoadmap, HuebscherMcCann2008AutonomicSurvey}. In CI/CD pipelines, authority is \emph{socially mediated}: multiple human stakeholders (developers, reviewers, release managers) share governance responsibility, and the ``managed element'' spans organizational processes, not just technical infrastructure. An agent that modifies a deployment threshold is not simply executing a control loop; it is intervening in a socio-technical governance structure. The data-plane/control-plane distinction captures this difference: data-plane actions resemble classical MAPE-K adaptations within a bounded execution context, while control-plane actions alter the governance structure, which MAPE-K does not capture. As capabilities expand, pipeline architectures should make authority boundaries explicit, distinguishing advisory, semi-autonomous, and control-plane roles. 

An important open question concerns autonomy escalation. As agents move from data-plane interventions to control-plane modifications, small expansions of authority may compound over time. For example, an agent initially permitted to generate patches may later be allowed to modify workflow definitions or deployment thresholds. Each incremental extension appears locally justified, yet collectively these changes may shift effective release governance from human-centered review toward agent-mediated control. Understanding how such escalation occurs, and how it can be bounded without eliminating useful automation, represents a critical design and policy challenge.


A further concern is adversarial robustness. Agents that generate workflow configurations or modify pipeline artifacts introduce a new attack surface for supply chain compromise. An agent-proposed change that passes CI checks but introduces a subtle vulnerability may be more difficult to detect than a conventional malicious commit, because reviewers may place unwarranted trust in agent-generated artifacts \cite{mittal2025leveraging}. Similarly, prompt injection or manipulation of the signals an agent observes (e.g., crafted log messages or error outputs) could steer remediation behavior in unintended directions. These risks are amplified at the control plane, where compromised modifications can affect release governance at an organizational level. Security implications of agentic CI/CD therefore extend beyond the confidentiality and integrity of individual artifacts to the trustworthiness of the delivery process itself.

\subsection{Summary}
Across the systems we examined, a dominant pattern emerges: semi-autonomous agents embedded within human-governed workflows with bounded authority and constrained action spaces. We now formalize these patterns as explicit observations.

\section{Core Observations}

Analysis of research prototypes and industrial systems reveals three consistent patterns, which we present as conceptual claims grounded in observed system behavior that characterize the current state of agentic CI/CD.

\subsection{Observation 1: Constrained Autonomy Dominates}

Although many systems are described as agentic, most operate under tightly bounded autonomy. Research prototypes from Repairnator through RepairAgent and ChatRepair generate patches but defer integration decisions to human reviewers \cite{MonperrusEtAl2019RepairnatorUbiquity, BaudryEtAl2021SoftwareRepairRobot, RepairAgent2025, ChatRepair2024}. Industrial tools across multiple vendors similarly route generated changes through pull request or merge request workflows \cite{GitHubAgenticCI2026, GitLabDuoFixPipelineDocs, NxSelfHealingCIDocs}. 

In practice, these systems exercise delegated authority within predefined action spaces rather than independent control over pipeline orchestration or deployment strategy. As noted earlier, no publicly documented system delegates control-plane authority without human approval. The dominant pattern is semi-autonomous assistance embedded within existing governance structures, mirroring earlier transitions from manual release engineering to automated CI/CD workflows, where automation sped execution without immediately displacing human release authority \cite{Shahin2017CIDCDAccess, AdamsMcIntosh2016ReleaseEngineering}. Agentic CI/CD appears to be following a similar incremental trajectory, which suggests that the current bounded-autonomy equilibrium is not accidental but reflects a recurring pattern in how organizations absorb automation into governance-sensitive processes.

\subsection{Observation 2: Governance Carries the Safety Burden}

Safety and accountability in current agentic CI/CD systems are achieved primarily through external governance mechanisms (e.g., pull request approval workflows, repository permissions, and policy gates) rather than intrinsic agent guarantees. None of the reviewed systems uses formal verification of agent behavior or built-in safety constraints independent of the surrounding infrastructure.

This reliance on external governance has a critical implication: as agent capabilities expand to include broader planning and artifact modification \cite{Hou2024LLMsForSESLR, LiuEtAl2024LLMAgentsSESurvey}, the safety model depends entirely on the strength and coverage of surrounding controls. Explicit reasoning about authority boundaries and escalation paths is largely absent from current research and industrial documentation, leaving the governance layer as an unexamined single point of trust.

This governance-first safety model echoes patterns observed in DevSecOps integration, where security automation is typically embedded within policy-constrained workflows rather than granted unconstrained authority \cite{anugula2025autoguard, MishraOtaiwi2020DevOpsSLR}. The extension of similar constraints to agentic CI/CD suggests that autonomy expansion will likely remain mediated by compliance and risk management requirements.

\subsection{Observation 3: Evaluation Lags Behind Deployment}

Many industrial platforms now offer agentic CI/CD features \cite{GitHubAgenticCI2026}, yet systematic evaluation methods for these systems remain underdeveloped. Traditional CI/CD metrics emphasized in prior surveys, such as build duration, deployment frequency, and failure rate \cite{MishraOtaiwi2020DevOpsSLR, Shahin2017CIDCDAccess, FitzgeraldStol2017ContinuousSE}, were designed for deterministic automation and do not capture decision quality, constraint adherence, escalation behavior, or long-term governance impact. Existing AIOps and DevOps evaluation frameworks primarily assess anomaly detection accuracy and operational performance gains \cite{Notaro2021AIOpsSurvey, MishraOtaiwi2020DevOpsSLR}, while recent empirical work highlights that generative AI systems lack clear correctness criteria and are difficult to evaluate in a deterministic manner~\cite{almulla2025understanding}, making such approaches inadequate for agentic systems that autonomously generate and execute actions on repository artifacts or deployments.

Evaluation must move beyond task-level correctness toward system-level assessment. An agent may successfully repair a failing build while increasing reviewer burden, triggering unstable rollback cascades, or subtly reshaping governance practices. Such socio-technical effects are not captured in conventional CI/CD benchmarking approaches, and no evaluation framework currently accounts for them in a standardized manner.

These three observations converge on a single conclusion. Constrained autonomy persists (Obs. 1) because governance mechanisms carry the safety burden (Obs. 2), while the lack of evaluation frameworks (Obs. 3) prevents systematic validation of alternative designs. Together, these constraints show that agentic CI/CD is not primarily a model-capability question, but a reallocation of decision authority within socio-technical delivery systems. Clarifying how authority is bounded, evaluated, and coordinated is therefore the central research challenge.

\section{Research Agenda}

We outline four research directions, ordered by urgency. Control-plane safety is the most pressing because it addresses the highest-risk frontier of capability expansion. The remaining directions provide the conceptual, empirical, and organizational foundations needed to support safe delegation.

\subsection{Control-Plane Safety and Governance Mechanisms}
\label{sec:agenda1}
As noted earlier, current systems avoid control-plane delegation. Expansion into control-plane modification introduces different risks, as such changes affect governance rather than individual artifacts. Adversarial risks are particularly critical at the control plane, where a single compromised modification can affect subsequent releases.

Research is needed on CI/CD-specific safety mechanisms, including policy-aware agents, constrained action synthesis, and automated rollback strategies with formal guarantees. The MAPE-K model offers a basis for feedback-loop design~\cite{Kephart2003AutonomicVision, Cheng2009SelfAdaptiveRoadmap}, but must be extended to account for socially mediated authority and multi-stakeholder governance in CI/CD pipelines.

\subsection{Formalizing Autonomy Boundaries}
\label{sec:agenda2}
Current systems implement bounded autonomy in an ad hoc manner, typically relying on pull request workflows and repository permissions to constrain agent behavior. A principled framework is needed to specify and verify autonomy boundaries in CI/CD pipelines. Such a framework could build on access control, policy specification, and runtime monitoring~\cite{HuebscherMcCann2008AutonomicSurvey}, extending these ideas to represent graded delegated authority rather than binary permissions. Recent work on LLM agent observability~\cite{dong2024agentops} suggests that runtime monitoring of agent decisions is feasible and could enable formal boundary enforcement.

One practical direction is to model decision authority, action scopes, and escalation paths explicitly within pipeline architectures. Such models would treat autonomy as a configurable design parameter rather than an emergent side effect. The data-plane/control-plane distinction we propose offers a starting vocabulary for such specifications.

\subsection{Evaluation Frameworks for Agentic CI/CD}
\label{sec:agenda3}
Existing CI/CD metrics overlook decision quality, constraint adherence, and governance impact, and the field lacks benchmarks for comparing autonomy levels and trade-offs. Recent benchmarks such as SWE-bench~\cite{swebench} evaluate LLM-based agents on issue resolution via patch generation, capturing primarily data-plane capabilities. However, they do not capture control-plane aspects such as policy constraints, approval workflows, or governance decisions, motivating CI/CD-specific evaluation frameworks.

Future work should develop evaluation environments that capture realistic pipeline workflows, approval processes, and failure scenarios. Related efforts exist in adjacent domains: AIOpsLab evaluates AI agents in cloud operations~\cite{chen2025aiopslab}, ITBench offers diverse real-world IT automation tasks for agent evaluation~\cite{jha2025itbench}, and AIDev analyzes agent-authored pull requests in development workflows~\cite{AIDev}. However, no comparable standardized environment exists for CI/CD pipelines that captures approval gates, policy constraints, control-plane changes, and human overrides.


Evaluation criteria should include stability, transparency, human override frequency, and long-term effects. A practical entry point is a minimal protocol separating remediation effectiveness from governance and stability costs, drawing on CI/CD and DevOps perspectives~\cite{Shahin2017CIDCDAccess, MishraOtaiwi2020DevOpsSLR} and AIOps failure-management concerns~\cite{Notaro2021AIOpsSurvey}. Benchmark suites for agentic CI/CD should report: remediation success (fraction resolved, time-to-repair under fixed budgets), governance impact (PR iterations, reviewer interventions, override frequency), stability and regression (recurrence and agent-induced regressions), and policy adherence (violations of scoped action constraints, especially control-plane actions). Such reporting enables comparison of bounded-autonomy designs without assuming full autonomy as the target outcome.

\subsection{Human--Agent Coordination Models}


Current systems rely heavily on human-in-the-loop approval, yet little research examines how developers interpret or override such decisions.

During incident response, ambiguous authority between an agent and an on-call engineer can delay mitigation or produce conflicting actions. Research is needed on interaction models that clarify responsibility boundaries and prevent ambiguous authority in such situations.
Empirical datasets such as AIDev~\cite{AIDev} provide a foundation for studying agent-developer interaction, though CI/CD-specific datasets remain absent.

Advancing agentic CI/CD requires progress across all four directions, with control-plane safety as the enabling prerequisite. Without governance mechanisms that can bound control-plane authority, broader autonomy delegation risks undermining the release discipline that CI/CD was designed to enforce~\cite{FitzgeraldStol2017ContinuousSE, Humble2010ContinuousDelivery}.

\section{Limitations}
Our discussion relies partly on publicly available industrial documentation, which may be incomplete or marketing-influenced, so we prioritize observable properties over performance claims. 
The space is evolving quickly and public evidence on outcomes remains limited, so we emphasize authority and governance over long-term effectiveness.

\section{Conclusion}

Agentic CI/CD is best characterized as \emph{bounded autonomy}, where agents operate within human-governed approval structures rather than exercising independent control. Across systems, we observe three patterns: constrained autonomy, safety enforced through external governance, and a gap between deployment and evaluation. The \emph{data-plane/control-plane} distinction clarifies the stakes: current systems remain safe by confining agents to the data plane, but this boundary will be increasingly tested. \emph{Authority transfer} is the central challenge, with control-plane safety and governance as immediate priorities, followed by autonomy boundaries, evaluation frameworks, and human--agent coordination. Treating authority transfer as a first-class design problem is essential for safe and accountable agentic CI/CD systems.

\begin{acks}
Funding: We acknowledge the support of the Natural Sciences and Engineering Research Council of Canada (NSERC): [RGPIN-2021-03969].
\end{acks}

\balance
\bibliographystyle{ACM-Reference-Format}
\bibliography{references}
\end{document}